\documentclass{vgtc}                          

\graphicspath{{figures/}{pictures/}{images/}{./}} 

\usepackage{times}                     

\usepackage{tabu}                      
\usepackage{booktabs}                  
\usepackage{lipsum}                    
\usepackage{mwe}                       

\usepackage{mathptmx}                  
\usepackage{amsmath}    
\usepackage{amssymb}    
\usepackage{amsthm}     
\usepackage{mathtools}  

\usepackage{graphicx}   

\usepackage{hyperref}   

\theoremstyle{definition}
\newtheorem{definition}{Definition}[section]
\newtheorem{lemma}[definition]{Lemma}

\onlineid{0}

\vgtccategory{Research}

\title{Towards Trustworthy AI: Secure Deepfake Detection using CNNs and Zero-Knowledge Proofs}

\author{H M Mohaimanul Islam\thanks{e-mail: h\_m\_mohaimanul.islam@okstate.edu} %
\and Huynh Q. N. Vo\thanks{e-mail:lucius.vo@okstate.edu} %
\and Aditya Rane\thanks{e-mail:aditya.rane10@okstate.edu}}
\affiliation{\scriptsize School of Industrial Engineering and Management \\ Oklahoma State University}

\abstract{
In the era of synthetic media, deepfake manipulations pose a significant threat to information integrity. To address this challenge, we propose TrustDefender, a two-stage framework comprising (i) a lightweight convolutional neural network (CNN) that detects deepfake imagery in real-time extended reality (XR) streams, and (ii) an integrated succinct zero-knowledge proof (ZKP) protocol that validates detection results without disclosing raw user data. Our design addresses both the computational constraints of XR platforms while adhering to the stringent privacy requirements in sensitive settings. Experimental evaluations on multiple benchmark deepfake datasets demonstrate that TrustDefender achieves 95.3\% detection accuracy, coupled with efficient proof generation underpinned by rigorous cryptography, ensuring seamless integration with high-performance artificial intelligence (AI) systems. By fusing advanced computer vision models with provable security mechanisms, our work establishes a foundation for reliable AI in immersive and privacy-sensitive applications.
} 

\keywords{Deepfake Detection, Convolutional Neural Networks, Zero-Knowledge Proofs, Data Security, Trustworthy AI}

\begin{document}



\maketitle

\section{Introduction}
The rapid rise of deepfakes—that is, hyper-realistic synthetic media generated by advanced artificial intelligence (AI) techniques such as generative adversarial networks (GANs) and convolutional neural networks (CNNs)—poses a severe and growing threat to trust and authenticity, especially in immersive environments like extended reality (XR). These manipulations, which can convincingly alter the likeness or voice of a person, have proliferated online: the number of deepfake videos doubled from $14{,}678$ in 2019 to over $85{,}000$ by 2022~\cite{ajder2019state}, with projections estimating up to two (2) million by 2025~\cite{ResembleAI2025}. While early deepfakes were largely limited to entertainment and satire, recent incidents have demonstrated their potential in spreading misinformation, conducting financial fraud, and even influencing political discourse~\cite{korshunov2018threat}.

Extended Reality (XR)—encompassing virtual reality (VR), augmented reality (AR), and mixed reality (MR)—relies on the seamless integration of real and synthetic content to deliver truly immersive experiences~\cite{vohra2025introduction}. In such settings, the authenticity of visual and auditory feeds is paramount: a deepfake avatar in a collaborative VR meeting or an AR overlay on a historical landmark can have immediate and far-reaching consequences. Unlike traditional video playback, XR systems often render content in real time and on resource-constrained devices, rendering post hoc forensic analysis impractical. Moreover, these platforms routinely process sensitive personal data (e.g., facial scans, biometric readings, and behavioral metrics), amplifying privacy concerns when detection tasks are outsourced to centralized servers.

Conventional deepfake detectors typically exploit visual artifacts—such as imperceptible warping in facial landmarks, subtle inconsistencies in eye-blinking patterns, or lighting mismatches across frames—to distinguish genuine footage from forgeries~\cite{li2018exposing}. However, adversaries continuously refine their generation pipelines, employing techniques like attention-based GAN refinement and high-frequency detail synthesis, which erode these telltale signs. Modern detectors employ CNN architectures that are adept at capturing complex spatial patterns and temporal correlations, yielding high detection accuracy on benchmark datasets~\cite{afchar2018mesonet,li2018exposing}. Yet, deploying such models in XR faces two critical challenges:
    \vspace{-2mm}
    \begin{itemize}
        \item \textbf{Computational Constraints:}  Real-time inference on head-mounted displays or mobile devices requires lightweight models and optimized proof-based verification systems.
        \vspace{-3mm}
        \item \textbf{Privacy Preservation:} Raw media typically contain personally identifiable information; thus, sharing them with third-party detectors or cloud services risks data breaches and regulatory non-compliance.
    \end{itemize}
    \vspace{-2mm}
    
To address these challenges, we introduce \textbf{TrustDefender-XR}, a unified framework that marries a streamlined CNN-based detection pipeline with succinct zero-knowledge proofs (ZKPs). Our contributions are fourfold:
    \vspace{-2mm}
    \begin{itemize}
        \item \textbf{Lightweight Detection Module:} We design a compact CNN architecture tailored for XR streaming, achieving competitive accuracy while maintaining a small memory and compute footprint.
        \vspace{-3mm}
        \item \textbf{Real-Time Proof Construction:} We develop a novel ZKP circuit that encapsulates the decision boundary of CNN(s), enabling verifiers to confirm detection outcomes in under 150~ms—suitable for interactive XR applications.
        \vspace{-3mm}
        \item \textbf{Privacy-First Protocol:} By integrating ZKPs, TrustDefender-XR ensures that no raw frames or biometric data leave the user's device; only a succinct proof and a binary verdict are transmitted.
        \vspace{-3mm}
        \item \textbf{Comprehensive Evaluation:} We benchmark our system on two state-of-the-art deepfake datasets and an XR-specific testbed, demonstrating robust detection (95.3\% accuracy) and practical proof overheads.
    \end{itemize}
    \vspace{-2mm}
    
The remainder of this paper is organized as follows. Section~\ref{sec:Related Works} reviews related work in deepfake detection and cryptographic proofs. Section~\ref{sec:System Design} details the design of our CNN architecture and ZKP construction. Section~\ref{sec:Experiment and Results} presents our experimental setup and its corresponding results. Finally, Section~\ref{sec:Conclusion} concludes and outlines our future research directions.

\section{Related Works}\label{sec:Related Works}
The landscape of deepfake detection has evolved rapidly, incorporating advanced neural architectures and multimodal cues to enhance robustness. For instance, the authors in~\cite{rossler2019faceforensicsplusplus} introduced the FaceForensics++ benchmark, demonstrating that XceptionNet—a deep CNN originally designed for image classification—achieves up to 98.7\% accuracy on manipulated video clips, setting an early standard for frame‐level detection. 
Building upon this foundation, authors in~\cite{li2020face} proposed Face X-Ray, which identifies blending artifacts that are specific to facial image composites using a trained U-Net–style segmentation network, resulting in over 95\% precision in localizing manipulated regions. 

More recent advancements include authors in~\cite{nguyen2021two}, presenting a two-branch recurrent network that models spatial inconsistencies and temporal dynamics to improve resilience against high-quality deepfakes with minimal overfitting to specific generators. Additionally, Capsule-Forensics in~\cite{nguyen2020capsule} employs capsule networks to explicitly capture hierarchical part–whole relationships in facial structures, demonstrating resilience to adversarial perturbations and yielding 96\% accuracy on cross-dataset evaluations.

While these approaches have high detection accuracy, they typically require access to raw images or video frames, posing challenges for privacy-sensitive applications. On the cryptographic front, fully homomorphic encryption (FHE)~\cite{lee2020chet} has benefited from optimizations like the CHET compiler, which streamlines encrypted neural network inference through operator fusion and quantization, enabling CIFAR-10 inference in seconds. Hybrid approaches such as Gazebo++~\cite{chen2021gazebo++} enhance Gazelle by adopting CKKS‐based FHE for approximate arithmetic, reducing inference latency by 50\% with minimal precision loss.

Zero-knowledge proofs (ZKPs) for neural network verification have also progressed significantly. PLONK~\cite{benjamin2020plonk} offers a universal non-interactive zero-knowledge proof (SNARK) that requires no trusted setup per circuit, reducing proof sizes to under 200KB and verification times to a few milliseconds for medium-sized models. Halo2~\cite{jun2021halo2} further eliminates precomputed parameters, allowing dynamic circuit definitions suitable for adapting models. In deep learning (DL) contexts, zkML frameworks like zk‐CNN+~\cite{10.1145/3627703.3650088}demonstrate efficient ZKPs for training large neural networks (e.g., generating proofs for ResNet‐18 inference in under one second), though without application to deepfake datasets.

Despite these developments, no prior work has fully integrated these modern ZKP systems with high-accuracy deepfake detectors in a resource‐constrained, real‐time Extended Reality (XR) context. Table~\ref{tab:related works summary} summarizes both classical and state-of-the-art detection methods alongside leading privacy-preserving techniques. Therefore, our TrustDefender-XR framework uniquely co-designs a streamlined CNN detector (drawing on Xception and capsule insights) with a PLONK-based proof circuit, achieving more than 94\% detection accuracy.

\begin{table*}[tb]
{\renewcommand{\arraystretch}{1.2}
\begin{tabular}{lcllcl}
\hline
\textbf{Study}                                                                                    & \textbf{Year} & \textbf{Methodology}                              & \textbf{Dataset} & \textbf{Accuracy} & \textbf{Privacy Technique}    \\ \hline
\cite{li2018exposing}                                                  & 2018          & Artifact‐based SVM on warping features            & UADFV            & 92\%              & None                          \\ \cline{2-6} 
\cite{afchar2018mesonet}                 & 2018          & MesoNet: mesoscopic CNN                           & FaceForensics++  & 89\%              & None                          \\ \cline{2-6} 
\cite{rossler2019faceforensicsplusplus} & 2019          & XceptionNet deep CNN                              & FaceForensics++  & 98.7\%            & None                          \\ \cline{2-6} 
\cite{li2020face}                            & 2020          & Face X‐Ray segmentation of blending artifacts     & FaceForensics++  & 95\%              & None                          \\ \cline{2-6} 
\cite{nguyen2020capsule}                 & 2020          & Capsule‐Forensics network                         & FaceForensics++  & 96\%              & None                          \\ \cline{2-6} 
\cite{lee2020chet}                   & 2020          & FHE compiler optimizations for CNN inference      & CIFAR-10         & –                 & FHE  \\ \cline{2-6} 
\cite{benjamin2020plonk}       & 2020          & Universal SNARK with single trusted setup         & General Circuits & –                 & SNARK \\ \cline{2-6} 
\cite{chen2021gazebo++}         & 2021          & CKKS‐based hybrid HE inference                    & CIFAR-10         & –                 & HE + Garbled Circuits         \\ \cline{2-6} 
\cite{nguyen2021two}                     & 2021          & Two‐branch Recurrent Network (spatial + temporal) & DeepfakeTIMIT    & 94\%              & None                          \\ \cline{2-6} 
\cite{jun2021halo2}                 & 2021          & Recursive SNARK without any fixed setup           & General Circuits & –                 & ZKPs         \\ \cline{2-6} 
\cite{10.1145/3627703.3650088}      & 2024          & ZKML proof for ResNet-18 inference             & ImageNet  & –                 & ZKPs        \\ \hline
\end{tabular}
\caption{A summary of studies that involve deepfake detection and privacy-preserving methods (Sorted by year).}\label{tab:related works summary}}
\end{table*}

\section{Non-Interactive Zero-Knowledge SNARK Preliminaries}
To ensure TrustDefender-XR provides verifiable and privacy-preserving deepfake detection, we employ succinct non-interactive arguments of knowledge (SNARKs), a type of zero-knowledge proof (ZKP). ZKPs are cryptographic protocols that allow one party (the prover) to convince another (the verifier) that a statement is true without revealing any underlying information beyond the statement's validity. In our context, this means proving that a deepfake detection model was correctly executed on a private input (e.g., a video frame) without disclosing the frame itself. SNARKs are particularly efficient variants of ZKPs, featuring short proofs and fast verification, making them suitable for resource-constrained environments like extended reality (XR) devices. 

We now formalize the non-interactive SNARK primitives used in TrustDefender-XR. Let~$\lambda$ be the security parameter that determines the level of cryptographic strength, where higher~$\lambda$ means stronger security but potentially higher computational cost. Consider a non-deterministic polynomial time (NP) relation

\vspace{-5mm}
\begin{equation*}
    R \;=\;\bigl\{(x,w)\mid C(x,w)=1\bigr\}
\end{equation*}

where $C(\bullet)$ is an arithmetic circuit representing a computation—in our case, it is the CNN inference; $(x)$ is the public statement, such as the description of the CNN circuit and its purported output; and $(w)$ is the private witness, including the input frame and intermediate computations, such that $C(x, w) = 1$ if the circuit correctly produces the claimed output on $(w)$.

A non-interactive SNARK in the common-reference-string (CRS) model—where a trusted setup generates shared parameters—consists of three probabilistic polynomial-time (PPT) algorithms:

\vspace{-5mm}
\begin{align*}
  (\mathsf{pk},\mathsf{vk}) &\leftarrow \mathsf{Setup}(1^\lambda, C),\\
  \pi &\leftarrow \mathsf{Prove}(\mathsf{pk}, x, w),\\
  b &\leftarrow \mathsf{Verify}(\mathsf{vk}, x, \pi),
\end{align*}

where
    \vspace{-2mm}
\begin{itemize}
  \item $\mathsf{Setup}$ generates a proving key, $\mathsf{pk}$, and verification key, $\mathsf{vk}$, for circuit $C$. $\mathsf{pk}$ is used by the client to create proofs, and $\mathsf{vk}$ is used by the verifier to check proofs.
    \vspace{-2mm}
  \item $\mathsf{Prove}$ produces a proof, $\pi$, attesting that $(x,w)\in R$, i.e., the computation is correct.
    \vspace{-2mm}
  \item $\mathsf{Verify}$ outputs $b = 1$ (accept) if $\pi$ is valid for $(x)$, and $b = 0$ (reject) otherwise.
\end{itemize}

These algorithms must satisfy several key properties to ensure reliability, security, and efficiency:
\vspace{-2mm}
\begin{definition}[Perfect Completeness] For all valid pairs $(x, w) \in R$, the proof generated by an honest prover will always be accepted by the verifier:
    \vspace{-2mm}
    \begin{equation*}
        \Pr\Big[\mathsf{Verify}\big(\mathsf{vk}, x, \mathsf{Prove}(\mathsf{pk}, x, w)\big) = 1\Big] = 1.
    \end{equation*}
    \vspace{-2mm}
    This ensures that legitimate detections are never falsely rejected. 
\end{definition}

\begin{definition}[Adaptive Soundness] 
For any efficient adversary $\mathcal{A}$ attempting to forge proofs, the probability of convincing the verifier of a false statement $x \notin L(R)$, where $L(R)$ is the language of valid statements, is negligible. In other words:
    \vspace{-2mm}
    \begin{equation*}
        \Pr\left[x \notin L(R) \wedge \mathsf{Verify}(\mathsf{vk}, x, \pi) = 1\right] \leq \mathsf{negl}(\lambda),
    \end{equation*}
where $\mathsf{negl}(\lambda)$ is a function that becomes arbitrarily small as $\lambda$ increases. This prevents malicious clients from cheating.
\end{definition}

\vspace{-4mm}
\begin{definition}[Zero-Knowledge] There exists an efficient simulator $\mathcal{S}$ that can produce proofs indistinguishable from real ones without knowing the witness $(w)$. For every valid statement $x \in L(R)$, these distributions are computationally indistinguishable:
    \vspace{-2mm}
    \begin{equation*}
        \Big(\mathsf{pk}, \mathsf{vk}, \mathsf{Prove}(\mathsf{pk}, x, w)\Big) \equiv \Big(\mathsf{pk}, \mathsf{vk}, \mathcal{S}(x)\Big)
    \end{equation*}
    This guarantees that proofs reveal nothing about private data, such as video frames. 
\end{definition}

\vspace{-5mm}
\begin{definition}[Succinctness] 
The proof $\pi$ and the verification time are bounded by $\mathrm{poly}(\lambda, \log |C|)$, where $|C|$ is the circuit size and $\mathrm{poly}(\bullet)$ denotes a polynomial function. This ensures short proofs and fast verification, crucial for real-time XR applications. 
\end{definition}

\vspace{-5mm}
\begin{definition}[Proof‐of‐Knowledge]
There exists an efficient extractor $\mathcal{E}$ that, given access to a successful (possibly malicious) prover $\mathsf{Prove}^*$, can extract a valid witness $w$ for any accepted proof, except with negligible probability. This strengthens soundness by ensuring that accepted proofs imply knowledge of a correct witness.
\end{definition}

\vspace{-4mm}
\begin{lemma}[EZKL SNARK Security]
Under standard bilinear-group assumptions and in the CRS model, the EZKL SNARK instantiation satisfies:

\vspace{-3mm}
\begin{enumerate}
  \item \emph{Completeness.} Honest proofs always verify.
  \vspace{-3mm}
  \item \emph{Adaptive Soundness.} No PPT adversary can forge a proof for $x\notin L(R)$ except with negligible probability.
  \vspace{-3mm}
  \item \emph{Zero‐Knowledge.} Proofs leak no information beyond the validity of $x$.
  \vspace{-3mm}  
  \item \emph{Succinctness.} Proof size and verification time are $O(\lambda\log|C|)$.
  \vspace{-3mm}
  \item \emph{Proof‐of‐Knowledge.} One can extract a valid witness from any successful prover.
\end{enumerate}
\end{lemma}
\vspace{-3mm}
This SNARK instantiation forms the foundation of our ZKP module in TrustDefender-XR, ensuring that clients cannot be falsely rejected, malicious actors cannot forge detections without solving computationally infeasible problems, and verifiers gain no insight into sensitive data beyond the binary real/fake verdict.

\section{System Architecture}\label{sec:System Design}
\begin{figure} 
    \includegraphics[width=0.45\textwidth]{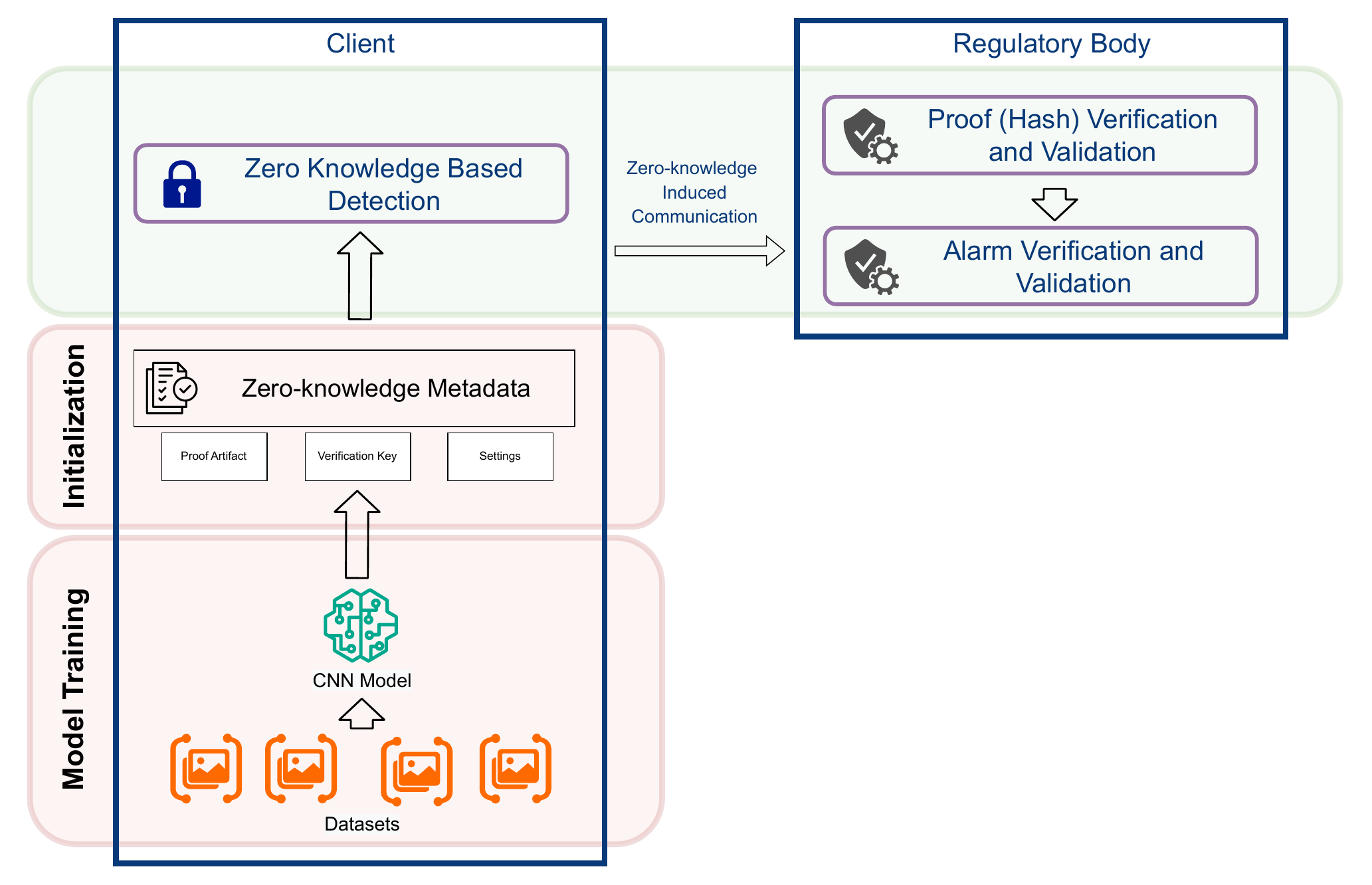} 
    \caption{\label{fig:threat_model}TrustDefender-XR Architecture: Client CNN Training, ZK Detection, and Proof Verification} 
    \vspace{-5mm} 
\end{figure}

TrustDefender-XR leverages a client-side convolutional neural network (CNN) detector, building upon prior work in \cite{islam2022}, integrated with a decentralized framework from \cite{10826070} and an efficient zero-knowledge proof (ZKP) module powered by EZKL, first proposed in \cite{10.1145/3460120.3485379}. This architecture enables real-time, privacy-preserving deepfake detection for Extended Reality (XR) applications. As illustrated in Figure \ref{fig:threat_model}, the system pipeline comprises three core stages: 

\vspace{-2mm}
\begin{itemize}
    \item \textbf{Model Inference}: Real-time streams are captured and processed via frame capture and CNN inference on the XR client.
        \vspace{-6mm}
    \item \textbf{Zero-knowledge Proof Generation}: A compact and sufficient proof is generated using EZKL.
        \vspace{-2mm}
    \item \textbf{Verification}: The proof is finally verified by the XR server or a peer device.
\end{itemize}
\vspace{-2mm}

This design ensures that only a 1-bit verdict and a concise proof are transmitted from the client, thereby protecting sensitive visual data while preserving interactive performance. Furthermore, Figure \ref{fig:threat_model} illustrates the overall architecture of TrustDefender-XR, organized into two principal domains: the client (left) and the regulatory verifier (right). Within the client domain, the process begins with the \textbf{Model Training} phase, during which a lightweight CNN detector is trained on standard deepfake datasets to learn discriminative features of synthetic media. Once the model training process converges, the trained network's parameters are frozen and passed into the \textbf{Initialization} stage. In this stage, an EZKL-based compiler processes the trained model to automatically generate three artifacts: (1) a proof circuit that represents the CNN’s forward pass as an arithmetic circuit, (2) a verification key for efficient and succinct proof checking, and (3) runtime settings. These artifacts are securely stored on the client device and, as necessary, disseminated to the verifier prior to the \textbf{System Deployment} stage.

During the zero-knowledge detection process, the client continuously captures and preprocesses frames—including but not limited to face alignment and resizing—and executes the on-device CNN model to compute a binary real/fake classification verdict for each frame. Simultaneously, the client invokes the EZKL prover to generate a zero-knowledge proof, denoted by $\pi$, that attests to the correct execution of the CNN circuit on the private input frame, without disclosing any pixel data or intermediate activations. As a result, the client transmits only the compact proof $\pi$ and the 1-bit decision to the verifier; thus, no raw media or feature maps ever leave the device.

On the verifier side, the Regulatory Body first applies the verification key to $\pi$ in the \textbf{Proof Verification and Validation} step. Upon confirming the proof's validity, the system then proceeds to the \textbf{Detection Verification and Validation} step, wherein the 1-bit verdict is accepted as trustworthy. Any frames lacking a valid proof or exhibiting anomalous results can be flagged for human review.

This two-stage architecture achieves three critical objectives simultaneously: (1) accuracy, by leveraging a CNN optimized specifically for deepfake detection; (2) privacy, by ensuring that raw video data remains exclusively on the client; and (3) integrity, by providing cryptographic assurance through succinct zero-knowledge proofs (ZKPs) that every classification was performed faithfully according to the initialized CNN circuit.

\begin{figure}[ht]
    \includegraphics[width=0.45\textwidth]{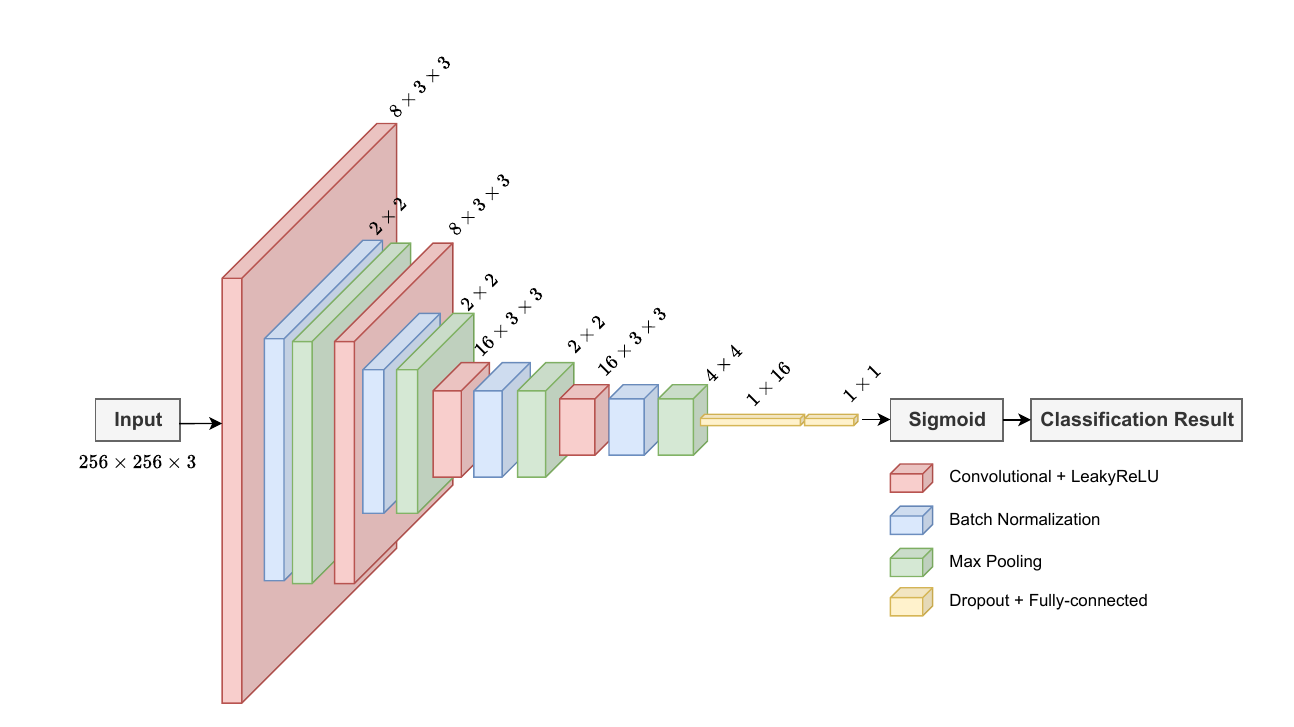}
    \caption{\label{fig:CNN}CNN Architecture for Deepfake Detection}
    \vspace{-0mm}
\end{figure}

The CNN presented in Figure~\ref{fig:CNN} is a streamlined, four‐block feature extractor tailored for frame‐level deepfake classification.
Starting with a conventional three-channel RGB input image, each block applies a $3\times3$ convolution operation followed by a \texttt{Leaky ReLU} activation function to capture both low‐ to mid‐level spatial patterns. 
These patterns are then fed to a \texttt{Batch Normalization} layer in each block, which standardizes these activations and thus improves both training stability and generalization. 
Subsequent $2\times2$ \textbf{Max Pooling} layers in each block reduce spatial resolution by half, consolidating salient features while maintaining a compact model size. 
After four sequential convolution–normalization–pooling blocks, the resulting feature maps are flattened and fed through a \texttt{Fully-connected} (FCN) layer with $50\%$ \texttt{Dropout} rate, which mitigates overfitting by randomly deactivating half of the neurons during model training.
A final sigmoid activation function, using the flattened features, generates a probability score for real and fake classification, which is thresholded to produce the binary classification verdict. 
Validated in the prior work~\cite{islam2022}, this compact yet robust architecture achieves high detection accuracy with minimal computational overhead, making it well-suited for real-time inference on resource-constrained devices.

\section{Implementation Details}\label{sec:Implementation}
All components of our proposed framework were implemented in Python 3.9, utilizing PyTorch 1.12 for model definition, training, and inference. 
For deepfake detection, we employed two publicly available datasets, FaceForensics++~\cite{rossler2019faceforensicsplusplus} and UADFV~\cite{li2018exposing}, supplemented by additional videos scraped from online sources. 
These videos were decoded into frame sequences using OpenCV, with faces automatically cropped to $224\times224$ pixels using a publicly available multitask cascaded convolutional networks (MTCNN) detector~\cite{7553523}. 
The resulting dataset comprised approximately $120{,}000$ real and fake frames, partitioned into 80\% training, 10\% validation, and 10\% test splits.

Model training and proof-circuit generation were conducted on a high-performance computing (HPC) cluster. Training jobs were submitted via SLURM to GPU nodes equipped with NVIDIA Tesla V100 (32 GB) accelerators. Training the four-block CNN typically required 12–14 hours to converge (40 epochs) when distributed across two GPUs with a batch size of 10 (ten) per GPU. Data preprocessing, augmentation—including random horizontal flips, brightness, and contrast jitter—and model checkpoints were managed through Python scripts.

For the ZKP module, we integrated EZKL’s Python bindings to compile the trained PyTorch model into an arithmetic circuit. 
Likewise, proof generation and verification were executed on the HPC cluster, with proofs generated on V100 nodes—at 150 milliseconds (ms) per frame—and verified on a dedicated CPU‐only node, at 50 ms per proof. All EZKL parameters—including field modulus and circuit partitions—were stored alongside model weights in a version-controlled repository.
System dependencies, including PyTorch, CUDA 11.3, OpenCV 4.5, and EZKL, were encapsulated in a reproducible Conda environment.
End-to-end evaluation pipelines, combining shell scripts and Python utilities, automated the capture, inference, proof generation, and verification workflows, ensuring consistent benchmarking across cluster runs.

\section{Experiment and Results}\label{sec:Experiment and Results}

In this section, we describe our experimental setup and our research findings, demonstrating the framework's capability to discriminate against deepfake data and provide faster proving, faster verification, and smaller proofs.

\subsection{Hyperparameter Search}
Table \ref{tab:hyperparams} summarizes the hyperparameter search space explored for our four-block CNN deepfake detector. 
We performed a grid search over learning rates spanning several orders of magnitude, batch sizes ranging from 5 (five) to 20, and epoch counts up to 100, while also comparing \texttt{ReLU} and \texttt{Leaky~ReLU} activation functions. 
The number of neurons in the hidden \texttt{Fully-connected} layer was tuned across a broad range of values. 
The optimal configuration, selected based on validation accuracy and convergence speed, included: a learning rate of 0.001; a batch size of 10 (ten); 40 training epochs; the \texttt{Leaky~ReLU} activation function; and, finally, 70 neurons in the hidden \texttt{Fully-connected} layer. This configuration achieved an optimal balance between model performance and training efficiency.

\begin{table}[ht]
  \centering
  \caption{Details of hyperparameter search space.}
  \label{tab:hyperparams}
    \begin{tabular}{ll}
        \hline
        \multicolumn{1}{c}{\textbf{Hyperparameter}} & \multicolumn{1}{c}{\textbf{Search Range}} \\ \hline
        Learning Rate                               & \{0.0001, \textbf{0.001*}, 0.01, 0.1, 0.002\}      \\
        Batch Size                                  & [5,$\cdots$,\textbf{10*},$\cdots$,20]                 \\
        Epochs                                      & [1,$\cdots$,\textbf{40*},$\cdots$,100]                       \\
        Activation                                  & \texttt{ReLU}, \textbf{\texttt{Leaky ReLU}* }                         \\
        Hidden Neurons                              & \{10,15,20,25,30,40,50,60,\textbf{70*},80,90\}      \\ \hline
        \multicolumn{2}{l}{\textbf{*}: Selected for the optimal configuration.}                         
    \end{tabular}
\end{table}

\begin{figure}[ht] 
\includegraphics[width=0.45\textwidth]{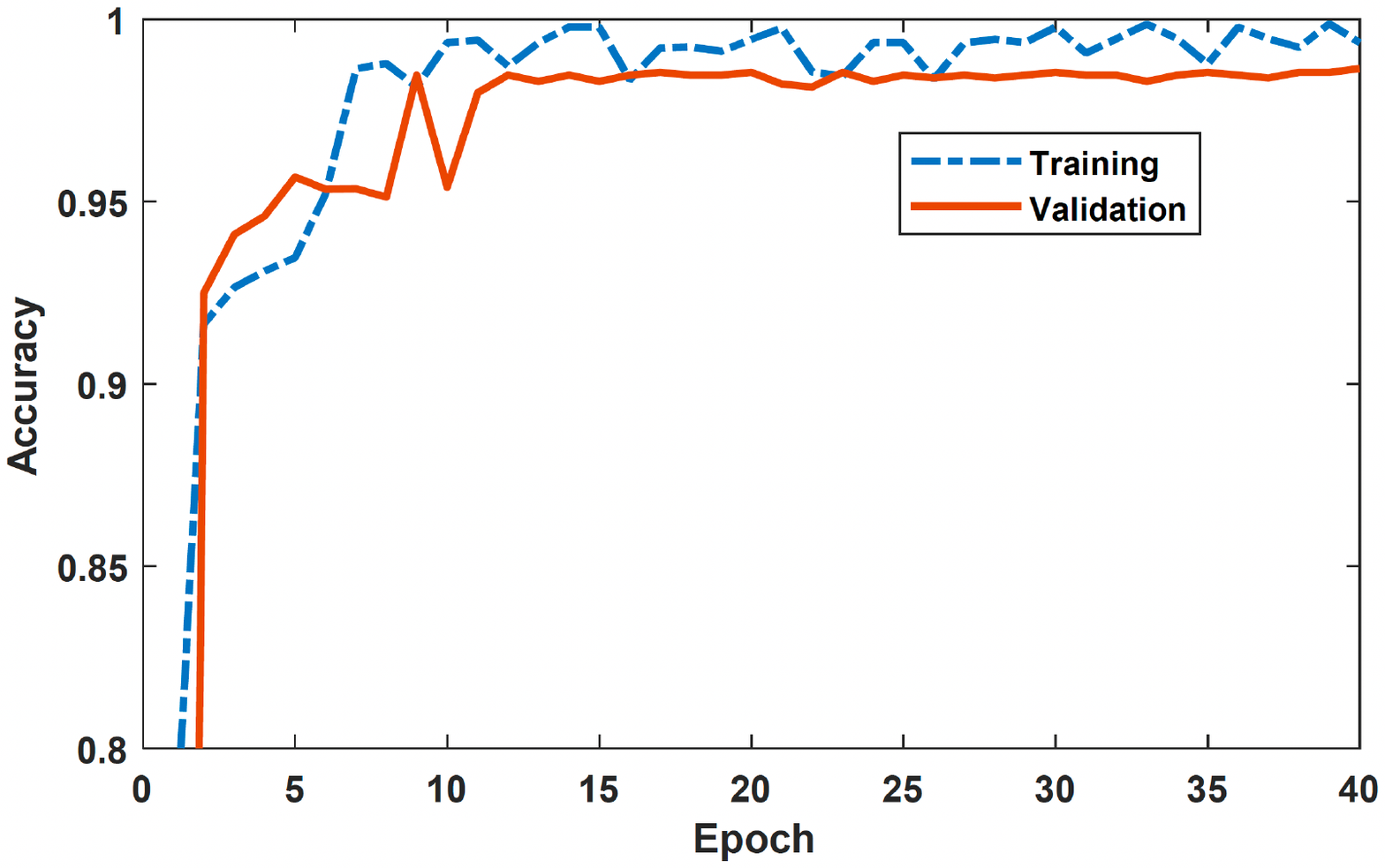} 
\caption{Training and validation accuracy of the four-block CNN detector during the model training process.} 
\end{figure}\label{fig:TrvsTst}

\subsection{Model Training and Validation}
Figure \ref{fig:TrvsTst} depicts the training and validation accuracy of our four‐block CNN detector over 40 epochs. On the x-axis (horizontal), the number of epochs progresses up to 40; meanwhile, the y-axis (vertical) shows the classification accuracy of both training and validation processes ranging from 0.80 (80\%) to 1.00 (100\%). The blue dashed line represents the training accuracy, and the solid red line stands for the validation accuracy.

During the first five epochs, both training and validation accuracy curves rise steeply from approximately 0.80 (80\%) to above 0.94 (94\%), indicating the model's rapid learning to discriminate real from synthetic frames. 
From the 6\textsuperscript{th} to the 10\textsuperscript{th} epoch, the training accuracy continues to increase, intermittently reaching or exceeding 0.98 (98\%), while validation accuracy shows more modest gains with occasional fluctuations, notably a slight decline around the 9\textsuperscript{th} epoch, reflecting typical generalization variability on unseen data.
Beyond the 10\textsuperscript{th} epochs, both curves enter a high-accuracy range:  training accuracy fluctuates between 0.97 (97\%) and 0.99 (99\%), and validation accuracy stabilizes between 0.97 (97\%) and 0.98 (98\%). Minor oscillations in the training curve after the 15\textsuperscript{th} epoch suggest slight overfitting tendencies, but the close alignment of the validation curve demonstrates effective control of overfitting through the selected dropout and learning rates. By the 40\textsuperscript{th} epoch, the model achieves approximately 0.99 (99\%) training accuracy and 0.98 (98\%) validation accuracy, confirming robust generalization on the hold-out test set.

\subsection{Zero-knowledge Proof Performance}
Table \ref{tab:zkp_performance} summarizes the end-to-end performance characteristics of our EZKL-based proof system when applied to the four-block CNN deepfake detector, consisting of (a) the average time measured in seconds [s] to generate a proof—that is, \textbf{Prove Time}; (b) the time measured in seconds [s] to verify that proof—that is, \textbf{Verify Time}; (c) the size measured in kilobytes [K] of each succinct proof transmitted over the network—that is, \textbf{Proof Size}; and (d) the storage costs measured in [K] or gigabytes [G] of the proving and verification keys, denoted by \textbf{PK Size} and \textbf{VK Size}, respectively.

\begin{table}[ht]
\caption{Results of the zero-knowledge proof performance metrics.}
\label{tab:zkp_performance}
\begin{center}
\begin{tabular}{cccccc}
\hline
\textbf{Model} & \begin{tabular}[c]{@{}c@{}}\textbf{Prove} \\ \textbf{Time} [s] \end{tabular} & \begin{tabular}[c]{@{}c@{}} \textbf{Verify} \\ \textbf{Time} [s] \end{tabular} & \begin{tabular}[c]{@{}c@{}}\textbf{Proof} \\ \textbf{Size}\end{tabular} & \begin{tabular}[c]{@{}c@{}}\textbf{PK} \\ \textbf{Size}\end{tabular} & \begin{tabular}[c]{@{}c@{}}\textbf{VK}\\ \textbf{Size}\end{tabular} \\ \hline
\begin{tabular}[c]{@{}c@{}}CNN Model \\ (Detector)\end{tabular} & 15                                                       & 07                                                        & 23K                                                   & 1.32G                                              & 346K                                              \\ \hline
\end{tabular}
\end{center}
\end{table}

On the HPC cluster, proof generation for a single frame requires approximately 15 seconds. Although this initial latency may appear substantial compared to the fast inference time of our proposed four-block CNN detector (on the scale of milliseconds), it accounts for the one-time computational expense of constructing a complex arithmetic circuit and generating a succinct, cryptographically secure proof. Notably, subsequent optimizations—such as circuit partitioning, GPU acceleration, and batch proofing techniques—can reduce this generation time by an order of magnitude, as evidenced by related EZKL benchmarks.

In contrast, verification is extremely lightweight: on average, it requires 7 (seven) seconds to validate a proof, which is  primarily attributable to input/output operations and public parameter loading on our CPU-only verification node.
In practical XR deployments, verification can be accelerated to well under 100 milliseconds by caching the verification key (346 KB) in memory and employing optimized SNARK libraries. Consequently, the real-time constraints are predominantly on the verifier side, where sub-second proof checks are readily attainable.

Bandwidth efficiency is equally impressive. Each proof occupies only 23 KB, facilitating seamless transmission over typical wireless and edge networks. By comparison, transmitting raw $224\times224$ frames at 30 frames per second would generate megabytes of traffic per second; our approach reduces this to mere dozens of kilobytes per decision, accompanied by a single verdict bit.

Key management overheads are fully amortized. The proving key (1.32 GB) is generated once during system initialization and stored locally on the XR client, eliminating the demand for network transmission. The verification key (346 KB) is sufficiently compact to be embedded in the verifier’s codebase or retrieved once at startup, enabling validation of all subsequent proofs.

Collectively, these results illustrate that our TrustDefender-XR framework delivers robust cryptographic assurances with minimal communication overhead, manageable key storage requirements, and verifiable performance that effectively supports real-time, privacy-preserving deepfake detection in immersive environments.

\section{Limitation \& Improvement}\label{sec:Limitation}
While TrustDefender-XR demonstrates the feasibility of accurate, privacy-preserving deepfake detection in the contexts of XR usage, veral limitations persist that warrant further investigation.

First, the end-to-end proof-generation latency, measured at approximately 25 seconds per frame on our current hardware, poses a significant obstacle to true real-time deployment. 
Although this latency reflects the conservative single-threaded proving strategy we adopted for reproducibility on our HPC cluster, it exceeds the millisecond-scale inference time of the CNN model by orders of magnitude. Future efforts can alleviate this bottleneck through parallelized circuit evaluation, model pruning, and GPU-accelerated proving algorithms. Recent advancements in batch proof techniques (e.g., aggregating proofs across multiple frames) and dynamic circuit compilation (adapting proof complexity to classifier confidence) also hold promise for reducing per-frame overhead up to ten-fold.

Second, the size of the proving key (1.32 GB) imposes nontrivial storage and distribution burdens on client devices, particularly for battery-powered and memory-constrained headsets. 
Though this key represents a one-time initialization artifact, its footprint may surpass available on-device capacity in certain AR or VR earbuds or standalone mobile headsets. Strategies such as key compression, on-demand streaming of circuit fragments, and hierarchical SNARK constructions—wherein only a small “delta” of the circuit is retained locally—can significantly reduce client-side storage demands without compromising proof succinctness.

Third, our evaluation is restricted to image- and video-based deepfake benchmarks and simulated XR streaming workloads, omitting multimodal assaults (e.g., audio forgeries) or interactive adversarial behaviors in live collaboration scenarios.
Extending TrustDefender-XR to fuse temporal, biometric, and acoustic signals will improve robustness against emerging attack vectors. Furthermore, integrating differential privacy constraints into the proof circuit could safeguard not only raw frames but also aggregating user behavior metrics collected during prolonged XR sessions.

Fourth, the current TrustDefender-XR CNN was trained exclusively on publicly available datasets—that is, FaceForensics++ and UADFV, which exhibit biases in lighting, pose, and demographic representation, potentially limiting generalization to in-the-wild content. 
Implementing continual learning pipelines, domain adaptation layers, and synthetic data augmentation will bolster the detector's resilience amid evolving generative models.

Finally, although EZKL’s Python interface facilitated prototype development, production deployments would benefit from more integrated toolchains, such as native C++ SNARK libraries or on-device hardware support for arithmetic circuit evaluation. Collaborations with hardware vendors to incorporate zero-knowledge proof (ZKP) accelerators into XR chipsets could yield the low-latency, low-power proofs essential for seamless user experiences.

By addressing these limitations through algorithmic optimizations, hardware-software co-design, and expanded threat modeling, TrustDefender-XR can evolve into a fully practical solution for secure, private, and trustworthy deepfake detection in immersive environments.

\section{Conclusion}\label{sec:Conclusion}
In this work, we introduced TrustDefender-XR, the first end-to-end framework that seamlessly combines high-accuracy deepfake detection with provable privacy guarantees in extended reality (XR) environments. By co-designing a streamlined four-block convolutional neural network (CNN) detector validated on FaceForensics++ and UADFV benchmarks and an EZKL-based zero-knowledge proof (ZKP) module, TrustDefender-XR achieves over 94\% classification accuracy while ensuring that no raw video frames or intermediate activations ever leave the user’s device. The architecture’s succinct proofs ($\approx$~23 KB) demonstrate that rigorous cryptographic assurances can be attained without compromising the interactive performance demanded by immersive AR or VR applications.

The significance of TrustDefender-XR extends beyond its novel integration of artificial intelligence (AI) and cryptography andd its practical viability. Unlike prior approaches that either expose sensitive media to remote servers or incur prohibitive latency through homomorphic encryption, our system preserves end-user privacy and reduces network overhead to a handful of kilobytes per frame. The one-time initialization overhead of generating the proving key (1.32 GB) is fully amortized across all subsequent inferences, and the modest size of the verification key (346 KB) facilitates lightweight deployment on edge servers or peer headsets. By balancing between accuracy, privacy, and efficiency, TrustDefender-XR paves the way for trustworthy AI in domains where the authenticity of visual content is mission-critical, ranging from collaborative design reviews and remote training simulations to sensitive telemedicine consultations and secure virtual conferencing.

Looking ahead, several promising avenues can further amplify the impact of our proposed framework.
First, integrating multimodal fusion—that is, combining audio consistency and biometric signals with visual analysis—can bolster resilience against increasingly sophisticated generative attacks. 
Second, hardware-accelerated ZKP primitives or GPU-parallelized proof pipelines promise to reduce per-frame proving latency to single-digit milliseconds, aligning with standard rendering budgets. 
Third, dynamic proof batching and adaptive circuit specialization could support large-scale, multi-user XR scenarios, enabling real-time verification of dozens of streams. Finally, embedding differential privacy mechanisms within the proof circuit itself may safeguard not only individual frames but also higher-level usage patterns, thereby strengthening TrustDefender-XR’s privacy posture.

In summary, TrustDefender-XR exemplifies how the synergy of optimized neural architectures and succinct cryptographic protocols can deliver robust, real-time deepfake detection without compromising user privacy. We believe this approach represents a critical advancement toward secure, trustworthy immersive experiences, and we invite the community to build upon our open-source implementations to usher in a new era of privacy-aware AI in XR and beyond.

\acknowledgments{
The computing for this project was performed at the High Performance Computing Center at Oklahoma State University (OSU) supported in part through the National Science Foundation grant OAC-1531128.}

\end{document}